\documentclass[aps,pre,amsmath,amssymb]{revtex4}
\usepackage{bm}
\usepackage{graphicx}
\begin{document}

\newcommand{\uu}[1]{\underline{#1}}
\newcommand{\pp}[1]{\phantom{#1}}
\newcommand{\be}{\begin{eqnarray}}
\newcommand{\ee}{\end{eqnarray}}
\newcommand{\ve}{\varepsilon}
\newcommand{\vs}{\varsigma}
\newcommand{\Tr}{{\,\rm Tr\,}}
\newcommand{\pol}{\frac{1}{2}}

\title{
Wavepacket of the Universe and its spreading}
\author{Marek Czachor$^{1,2}$ and Andrzej Posiewnik$^{3}$}
\affiliation{
$^1$ Katedra Fizyki Teoretycznej i Informatyki Kwantowej,
Politechnika Gda\'nska, 80-233 Gda\'nsk, Poland\\
$^2$ Centrum Leo Apostel (CLEA),
Vrije Universiteit Brussel, 1050 Brussels, Belgium,
\\
$^3$ Instytut Fizyki Teoretycznej i Astrofizyki,
Uniwersytet Gda\'nski, 80-952 Gda\'nsk, Poland\\
}

\begin{abstract}
Wavepackets in quantum mechanics spread and the Universe in cosmology expands. We discuss a formalism where the two effects can be unified. The basic assumption is that the Universe is determined by a unitarily evolving wavepacket defined on space-time. Space-time is static but the Universe is dynamic. Spreading analogous to expansion known from observational cosmology is obtained if one regards time evolution as a dynamical process determined by a variational principle employing Kolmogorov-Nagumo-R\'enyi averages. The formalism automatically leads to two types of ``time" parameters: $\tau$, with dimension of $x^0$, and dimensionless $\varepsilon=\ln \epsilon_\tau$, related to the form of diffeomorphism that defines the dynamics. There is no preferred time foliation, but effectively the dynamics leads to asymptotic concentration of the Universe on spacelike surfaces which propagate in space-time. The analysis is performed explicitly in $1+1$ dimensions, but the unitary evolution operator is brought to a form that makes generalizations to other dimensions and other fields quite natural.
\end{abstract}
\maketitle

\section{Introduction}

Normalizable wavepackets determine regions of space where quantum particles can be found. Such wavepackets spread due to their Schr\"odinger dynamics, so the regions expand with time. In cosmology, an analogous role is played by the size of the Universe --- its growth with time is described by Hubble's law. The two effects are universal, but apparently unrelated.

The goal of this paper is to consider a simple model of a Schr\"odinger dynamics that, in principle, might lead to a unifying framework for both phenomena. The case we discuss has been simplified to its extremes. We begin with $1+1$ dimensional empty Universe.
However, we believe that what we do is not entirely trivial and paves a way to rather obvious generalizations.

To begin with, we do not identify the dynamical Universe with dynamical space-time. Space-time is static, but the Universe is dynamic. This is possible, since what we regard as the Universe is, roughly speaking, a region of space-time associated with the support of the wavepacket. There is no Universe in those regions of space-time where the wavefunction is exactly zero. Moreover, in wavepackets such as Gaussians in space-time, the support of the wavepacket might include the whole of space-time, but nevertheless the ``effective size" of the Universe should not be infinite. What we expect is a measure of size analogous to a half-width of the wavepacket. The measure we take as the most natural one is the average value of an operator representing squared geodesic distance computed along spacelike directions. Our Universe diffuses in space-time.

Secondly, the evolution we propose leads to a dynamical ``localization of space-time" in neighborhoods of spacelike hypersurfaces. What it means is that our ``space" is not just a foliation of space-time into spacelike hypersurfaces (i.e. lines in $1+1$) parametrized by ``time". The ``space" has some thickness in timelike directions, but the dynamics shrinks this timelike thickness towards zero. The effect is compensated by spreading of the ``size of space" in spacelike directions. The two effects match each other in a way which guarantees conservation of norm of our wavefunction. This is how we represent the Hubble law. So, the Universe expands because the ``moment of now" becomes more and more concrete, and less and less fuzzy.

Now, what kind of space-time is the arena for our Universe? We decided to take a part of the Minkowski space that can be uniquely foliated by hyperbolas, so the support of the Universe is contained in one of the timelike cones. The choice of a future-pointing or a past-pointing cone is a matter of convention. We take the future cone $x_ax^a=x^2=s^2>0$, $x_0>0$, in order to avoid awkward-looking minuses in formulas, but the price we pay is that the Universe seems to evolve ``backward" in $x_0$ but forward in proper time $\tau$. In effect, the support of our Universe gets approximately localised on hyperbolas which asymptotically approach the light cone $s=0$. One can say that the proper time indeed flows in our model. This should be contrasted with the usual dynamics in space, which is equivalent to statics in space-time. In our model a distant past as well as a distant future with respect to ``now" literally do not exist in the deepest ontological sense. Interestingly, the evolution operator can be written as $e^{-i\varepsilon \hat\Phi}$, where $\hat\Phi$ is time-independent but $\varepsilon$ is a dimensionless parameter that for large $\tau$ becomes proportional to $\tau$, while in a distant past differs from $\tau$ in an essential way, a subtlety influencing possible interpretations of the origin of the model Universe. In addition to unitarity we thus also have conservation of energy, with $\tau$-independent Hamiltonian $\hat\Phi$.

As usual in quantum mechanics,  one can switch between Schr\"odinger and Heisenberg pictures. The Hubble law may be then represented in a form of a time-evolving operator of geodesic distance. This Heisenberg-Hubble equation is a departure point for less trivial generalizations, where the Hubble ``constant" evolves in proper time. The issue reduces to finding an appropriate one-parameter group of diffeomorphisms whose pull-back to the level of the wave function implies a Heisenberg picture dynamics of the geodesic position operator qualitatively agreeing with observational cosmology \cite{Weinberg}.

One such model naturally appears if one relates time evolution with an extremal entropy principle of the type discussed in the 1930s by Volterra \cite{Volterra}. For Shannon's entropy one gets an exponential expansion. Starting with R\'enyi $q$-entropies one finds a one-parameter family of possible expansions. The model that predicts a $\tau^{1/2}$ expansion of an early Universe, accompanied by a crossover to exponential expansion for later $\tau$s, occurs in the $q=2$ case. Since R\'enyi entropy of order $q=2$ is directly related to the correlation dimension, the extremal entropy principle is then interpretable as an ``extremal correlation dimension of time" principle, an issue intriguing in itself and worthy of further studies in the context of fractal structures of the Universe
\cite{Piet1,Piet2,Sylos,Balian,Scale2,Gaite,Bagla,Cordona}.

In final sections the unitary evolution operator is brought to a form which does not explicitly depend on dimensionality of the problem and emptiness of the Universe, and thus opens a way to higher dimensional generalizations.

\section{Universe associated with 1+1 dimensional space-time}

We first have to define what we mean by the Universe and its wave function. Let us begin with the Minkowski space of one time and one space dimensions. The future light-cone $V_+$ of some event $x^a=0$, i.e. $V_+=\{x^a\in \mathbb{R}^2;\,x_ax^a=x_0^2-x_1^2=s^2>0;\, x_0>0\}$ will play a role of a background space-time of the Universe. Now consider a square-integrable function $\psi(x^0,x^1)$, with the norm defined by
\be
\langle\psi|\psi\rangle
&=&
\int_{V_+}dx^0dx^1\,|\psi(x^0,x^1)|^2=\int_0^\infty ds\int_{-\infty}^\infty \frac{dx}{\sqrt{1+x_1^2/s^2}}\,\big|\psi(\sqrt{s^2+x_1^2},x^1)\big|^2.
\ee
Let us note that the integration is over the 1+1 dimensional volume. However, the intuition behind the construction is that the size of the Universe is related to the size of the wave-packet $\psi(\sqrt{s^2+x_1^2},x^1)$ measured with respect to the geodesic distance on the hyperbola $x_0^2-x_1^2=s^2$. An appropriate unitary dynamics should spread the wave-packet on the hyperbola, simultaneously maintaining the overall 1+1 dimensional norm. Yet another way of phrasing the basic intuition is that at certain stage of the dynamics of the Universe the wave-function should be well localized in $s$ around a given hyperbola, simultaneously being spread over the hyperbola in such a way that its average one-dimensional geodesic width should be comparable to the present-day size of our Universe. The fuzzyness of $s$ means that the notion of ``now" is smeared out as well, but in a present-day Universe this uncertainty of
``now" should be small, say of the Planck time scale.

Let us take an arbitrary fiducial point $X^a$ on the $s$ hyperbola, say with coordinates
\be
X^0 &=& s\cosh\Xi,\\
X^1 &=& s\sinh\Xi,
\ee
and an arbitrary point $x^a$ with coordinates
\be
x^0 &=& s\cosh(\Xi+\xi),\\
x^1 &=& s\sinh(\Xi+\xi),
\ee
where $s|\xi|$ is the geodesic distance between $x^a$ and $X^a$ evaluated along the hyperbola (yet another covariant definition is $X_ax^a/s^2=\cosh\xi$).
The two points satisfy the constraint
\be
x^2=x_ax^a=X^2=X_aX^a=s^2. \label{constr}
\ee
Changing $x^a$ we have to make sure that $X^a$ changes as well in a way which preserves the constraint (\ref{constr}). It is therefore perhaps better to speak of the fiducial field $X^a(x)=\sqrt{x^2}v^a$, where
\be
v^0 &=& \cosh\Xi,\\
v^1 &=& \sinh\Xi,
\ee
is the fiducial 4-velocity.
The Minkowski metric satisfies
\be
(dx^0)^2-(dx^1)^2=(ds)^2-s^2(d\xi)^2
\ee
and thus $a(s)=s$ is the Robertson-Walker scale factor while $s$ is the usual ``time" employed in cosmology \cite{Weinberg}.
Denote $\eta=s^2/2$ and
\be
\psi\big(s\cosh(\Xi+\xi),s\sinh(\Xi+\xi)\big)=f_v(\eta,\xi).
\ee
A change of the fiducial velocity $\Xi\to\Xi'$ is equivalent to a Lorentz transformation $v_a\to v'_a=\Lambda{_a}{^b}v_b$. The norm expressed in terms of $\xi$ and $\eta$ becomes
\be
\langle\psi|\psi\rangle
&=&
\int_0^\infty d\eta\int_{-\infty}^\infty d\xi\,\big|f_v(\eta,\xi)\big|^2.
\ee
In order to introduce a unitary dynamics $\psi\mapsto U_\tau\psi$ we consider a one-parameter family of diffeomorphisms $(\eta,\xi)\mapsto \phi_\tau(\eta,\xi)=(\eta_\tau,\xi_\tau)\in \mathbb{R}_+\times \mathbb{R}$ that will serve as a change of variables in the above integral.
We restrict $\phi_\tau$ to transformations do not changing ranges of integration, i.e. $0<\eta_\tau<\infty$, $-\infty<\xi_\tau<\infty$.
Then
\be
\langle\psi|\psi\rangle
&=&
\int_0^\infty d\eta_\tau\int_{-\infty}^\infty d\xi_\tau\,\big|f_v(\eta_\tau,\xi_\tau)\big|^2
\nonumber\\
&=&
\int_0^\infty d\eta\int_{-\infty}^\infty d\xi\,|J_\tau|\big|f_v\big(\phi_{\tau}(\eta,\xi)\big)\big|^2
\nonumber
\ee
where $J_\tau$ is the Jacobian. In this way we have arrived at the unitary representation
\be
U_\tau f_v(\eta,\xi) &=& \sqrt{|J_\tau|}f_v\big(\phi_{\tau}(\eta,\xi)\big)
\ee
of the one-parameter group in question. Returning to the original variables $x^a$ we obtain a representation $U_\tau\psi$. Our construction bears a similarity to some ideas known from unitary representations of groups defined in terms of quasi-invariant measures \cite{Mackey}, the Koopman-von Neumann representation of classical mechanics \cite{K,vN}, or the Dashen-Sharp-Goldin formulation of unitary representations of local currents \cite{DS,G}. On the other hand, however, we do not see any obvious links to wavefunctions defined on the superspace of different geometries, such as the classic formalisms of Wheeler-DeWitt \cite{P,W,dW} or Hartle-Hawking \cite{HH}.

\section{Dynamics (first attempt)}

We do not know why time flows, but the process seems to be related to changes in entropy.
Apparently, the first variational principle linking dynamics with entropy was proposed by Volterra in 1930s \cite{Volterra}, although the term ``entropy" was not explicitly used in this context. Volterra's principle involved abundances of species, but from a dynamical point of view it was applicable to any system of equations involving non-negative variables (such as our $s$ and $\eta$). A distinctive feature of Volterra Lagrangians is the presence of ``entropic" terms of the form $\dot q_j\ln\dot q_j$, where $q_j$ are configuration-space generalized coordinates. Let us consider a simple Volterra Lagrangian,
\be
L(q,\dot q,\tau)=-\sum_j \dot q_j\ln \dot q_j+\sum_j\dot q_j a_j,\label{Lagr}
\ee
where $a_j$  are $\tau$-dependent coefficients, supplemented by the constraint $\dot q_0+\dot q_1=C=\textrm{const}$, reducing the number of degrees of freedom to one. Integrated Euler-Lagrange equation reads
\be
\frac{\partial L}{\partial \dot q_0}
&=&
-\ln \dot q_0+\ln \dot q_1+a_0-a_1 =C_1\label{C_1},
\ee
where $C_1$ is a constant of motion and we have employed $\partial \dot q_1/\partial \dot q_0=-1$. In conclusion, $\dot q_1(\tau)=e^{C_1+ a_1-a_0}\dot q_0(\tau)$.
Note that $\dot q_j(\tau)$ are nonnegative by assumption, similarly to what one expects from $\eta_\tau$, so $q_j(\tau)$ are monotonic. 

The simplest nontrivial case of (\ref{C_1}) is $C_1=0$ and $a_j(\tau)=\lambda\tau j$, $j=0,1$, where $\lambda$ is a constant. Then, a
very similar derivation can be performed in a maximal-entropy thermodynamic formalism. The Lagrangian now plays a role of the Massieu function \cite{N} (a kind of free energy), and instead of solving Euler-Lagrange equations we look for its conditional extremum under the constraint that probabilities sum to 1.
The Massieu function involves a single Lagrange multiplier $\alpha$,
\be
E_{\rm S} &=&\sum_{j=0}^1 p_j\ln(1/p_j)+\alpha \sum_{j=0}^1 p_j+\tau\sum_{j=0}^1 \lambda j p_j,\\
\frac{\partial E_{\rm S}}{\partial p_0}
&=&
-\ln p_0-1+\alpha=0,\\
\frac{\partial E_{\rm S}}{\partial p_1}
&=&
-\ln p_1-1+\alpha+\lambda\tau=0.
\ee
Subtracting both equations we obtain
\be
\frac{\partial E_{\rm S}}{\partial p_0}-\frac{\partial E_{\rm S}}{\partial p_1}
&=&
-\ln p_0+\ln p_1-\lambda\tau=0
\ee
which is the same as (\ref{C_1}) with $C_1=0$,
and thus $p_1 = p_0 e^{\lambda\tau}$. The standard thermodynamic variational principle turns out to be a special case of the dynamical one. The final solution is
\be
p_0(\tau) &=& \frac{1}{e^{\lambda\tau}+1}=\dot q_0(\tau),\\
p_1(\tau) &=& \frac{e^{\lambda\tau}}{e^{\lambda\tau}+1}=\dot q_1(\tau),\\
q_0(\tau) &=& \frac{\lambda  \tau -\ln \left(1+e^{\lambda  \tau }\right)+\ln 2}{\lambda }+q_0(0).
\ee
$q_0(\tau)$ is strictly monotonic and thus invertible for all $\tau$. For $\lambda>0$ the probabilities following from the Volterra variational principle satisfy the asymptotics $p_0(-\infty)=1$, $p_1(-\infty)=0$, $p_0(+\infty)=0$, $p_1(+\infty)=1$. For $\lambda<0$ the roles of $p_0$ and $p_1$ are interchanged. Effectively, it is the product $\lambda\tau$ which determines the `arrow of time', with the two probabilities representing initial and final populations of the system in question.

The effective evolution parameter associated with the Volterra process satisfies 
\be
\lambda\tau=
\ln\big(p_1(\tau)/p_0(\tau)\big)
\ee
and thus one can {\it define\/} the evolution parameter in terms of the two populations as $\lambda\tau=\ln(p_1/p_0)$. 

Having in mind future generalization in terms of R\'enyi entropies, let us experiment with a simple exponential map associated with the Volterra process, $\eta_\tau=e^{\lambda \tau}\eta=(p_1/p_0)\eta$, $\xi_\tau=e^{-\lambda \tau}\xi$, $J_\tau=1$. It leads to the unitary transformation
\be
U_\tau f_v(\eta,\xi) &=& f_v(e^{\lambda \tau}\eta,e^{-\lambda \tau}\xi)
\nonumber\\
&=&
\psi\big(e^{\lambda \tau/2}\sqrt{2\eta}\cosh(\Xi+e^{-\lambda \tau}\xi),e^{\lambda \tau/2}\sqrt{2\eta}\sinh(\Xi+e^{-\lambda \tau}\xi)\big)
\nonumber\\
&=&
U_\tau \psi(x^0,x^1).
\ee
Let us make a remark that $\eta_\tau$ corresponds to the scale factor
\be
a_\tau=s_\tau=\sqrt{2\eta_\tau}=e^{\lambda \tau/2}\sqrt{2\eta}=e^{\lambda \tau/2}a_0
\ee
which resembles the inflation-phase dependence of scale on time.

Spreading of this wave packet can be illustrated in several ways. First of all, we introduce the operator of geodesic position
\be
\hat r_v f_v(\eta,\xi) &=& \sqrt{2\eta}\xi\, f_v(\eta,\xi),
\ee
or equivalently
\be
\hat r_v \psi(x^0,x^1) &=& \sqrt{x^2}
\Bigg(\,\textrm{arsinh}\frac{x^1}{\sqrt{x^2}}-\textrm{arsinh}\,v^1
\Bigg) \psi(x^0,x^1).
\ee
The size of the wavepacket is thus given by $R=\sqrt{\langle\psi|\hat r_v^2 \psi\rangle}$, so we can compute
\be
R_\tau^2
&=&
\langle U_\tau \psi|\hat r_v^2 U_\tau \psi\rangle
=
\int_0^\infty d\eta\int_{-\infty}^\infty d\xi\,2\eta\xi^2\big|U_\tau f_v(s,\xi)\big|^2
\nonumber\\
&=&
\int_0^\infty d\eta\int_{-\infty}^\infty d\xi\,2\eta\xi^2\big|f_v(e^{\lambda\tau}\eta,e^{-\lambda\tau}\xi)\big|^2
=
\int_0^\infty d\eta'\int_{-\infty}^\infty d\xi'\,e^{\lambda\tau}2\eta'\xi'^2\big|f_v(s',\xi')\big|^2
\nonumber\\
&=&
e^{\lambda\tau}
\langle \psi|\hat r_v^2 \psi\rangle.\label{14}
\ee
Spreading is here exponential, $R_\tau=e^{\lambda\tau/2}R_0$, which is the same rule as for the scale factor, so we get the usual formula
\be
\frac{1}{a_\tau}\frac{da_\tau}{d\tau}=\frac{1}{R_\tau}\frac{dR_\tau}{d\tau}
\ee
relating distance and scale. However, it must be stressed that in cosmology the derivative is over ``time" that would be typically identified with $s$, and not with our $\tau$.

An interesting alternative interpretation is possible if one interprets (\ref{14}) in terms of the Heisenberg picture. Indeed, what we have obtained is
equivalent to
\be
U_\tau^{\dag}\hat r_v^2 U_\tau =e^{\lambda\tau}\hat r_v^2
\ee
or
\be
U_\tau^{\dag}\hat r_v U_\tau =e^{\lambda\tau/2}\hat r_v.
\ee
A similar result is obtained for the ``proper time" operator
\be
\hat s f_v(\eta,\xi) &=& \sqrt{2\eta}\, f_v(\eta,\xi),\\
\hat s \psi(x^0,x^1) &=& \sqrt{x^2}\psi(x^0,x^1),\\
U_\tau^{\dag}\hat s\, U_\tau &=& e^{-\lambda\tau/2}\hat s.
\ee
The Hubble constant is in this simple example indeed a constant
\be
\frac{d}{d\tau}U_\tau^{\dag}\hat r_v U_\tau &=& \frac{\lambda}{2}U_\tau^{\dag}\hat r_v U_\tau=H_0 U_\tau^{\dag}\hat r_v U_\tau
= -i[U_\tau^{\dag}\hat r_v U_\tau,\hat\omega]
\ee
where $\hat\omega$ is the generator of $U_\tau$. Defining $\hat\omega_\tau=iU_\tau^\dag dU_\tau/d\tau$, we can write a general Heisenberg-Hubble equation
\be
H_\tau U_\tau^{\dag}\hat r_v U_\tau
&=& -i[U_\tau^{\dag}\hat r_v U_\tau,\hat\omega_\tau]
\ee
where $H_\tau a_\tau=da_\tau/d\tau$. This is the simplest equation linking metric tensor with the unitary dynamics.

In order to show the dynamics of probability density $|U_\tau\psi(x^0,x^1)|^2$ we have to reexpress the formulas directly at the level of $x^0$ and $x^1$.
This is simplest in the rest frame of the fiducial point, i.e. with $\Xi=0$, but even then the formula is rather cumbersome and counterintuitive,
\be
U_\tau \psi(x^0,x^1)
&=&
\psi\big(e^{\lambda \tau/2}s\cosh(e^{-\lambda \tau}\xi),e^{\lambda \tau/2}s\sinh(e^{-\lambda \tau}\xi)\big)
\nonumber\\
&=&
\psi\left(
e^{\lambda \tau/2}(x_+x_-)^{(1-e^{-\lambda \tau})/2}\frac{x_+^{e^{-\lambda \tau}}+x_-^{e^{-\lambda \tau}}}{2}
,
e^{\lambda \tau/2}(x_+x_-)^{(1-e^{-\lambda \tau})/2}\frac{x_+^{e^{-\lambda \tau}}-x_-^{e^{-\lambda \tau}}}{2}
\right)
\nonumber
\ee
where $x_\pm=x^0\pm x^1$. The next four figures show the dynamics of $|U_\tau\psi(x^0,x^1)|^2$ for a wavepacket which is initially well localized in space and time. So, in this picture, at $\tau=0$ the Universe is in superposition of various positions $x^1$ and times $x^0$, but one should bear in mind that $x^0$ is not the evolution parameter.
\begin{figure}
\includegraphics[width=12cm]{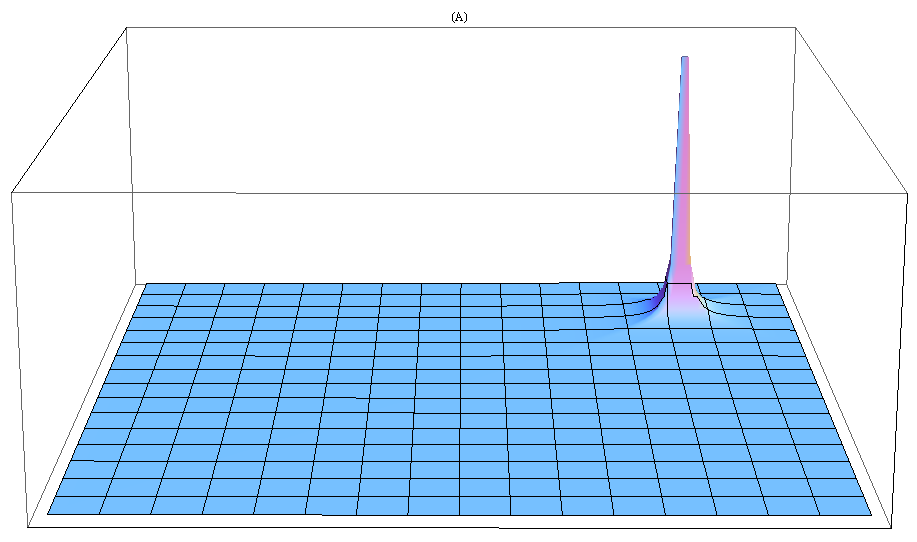}
\caption{The initial wavepacket at $\tau=0$. Its support defines the region in space-time occupied by the Universe. At this stage there is no preferred foliation of space-time into spacelike hypersurfaces.}
\end{figure}
\begin{figure}
\includegraphics[width=12cm]{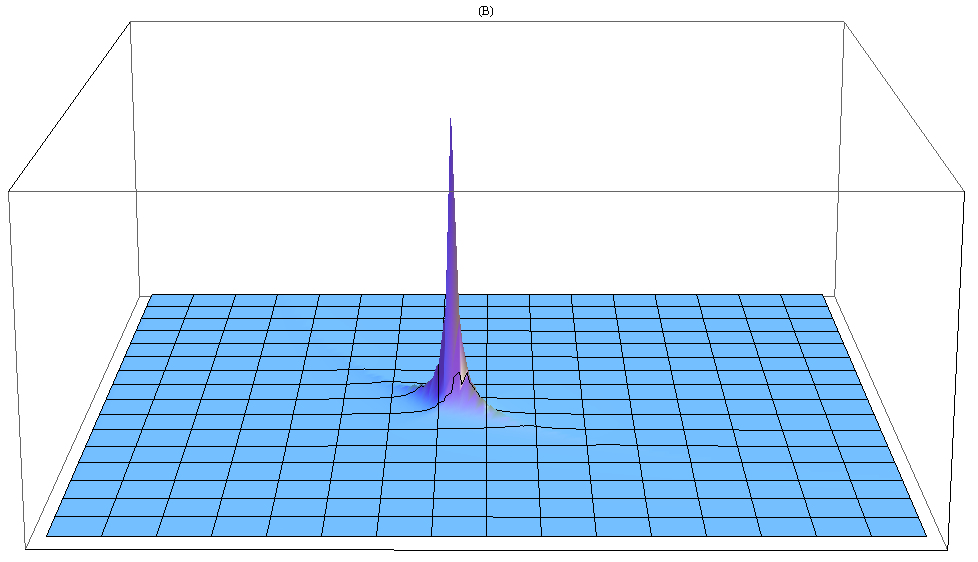}
\caption{Now $e^{-\lambda\tau}=0.3$. The universe starts to evolve in space-time and shifts towards the light cone, which forms the boundary of available background space-time. Note that from the point of view of $x^0$ the evolution seems to occur backwards in time. The choice of future and past is a matter of convention, since $x^0$ is not the evolution parameter but a component of space-time position operator.}
\end{figure}
\begin{figure}
\includegraphics[width=12cm]{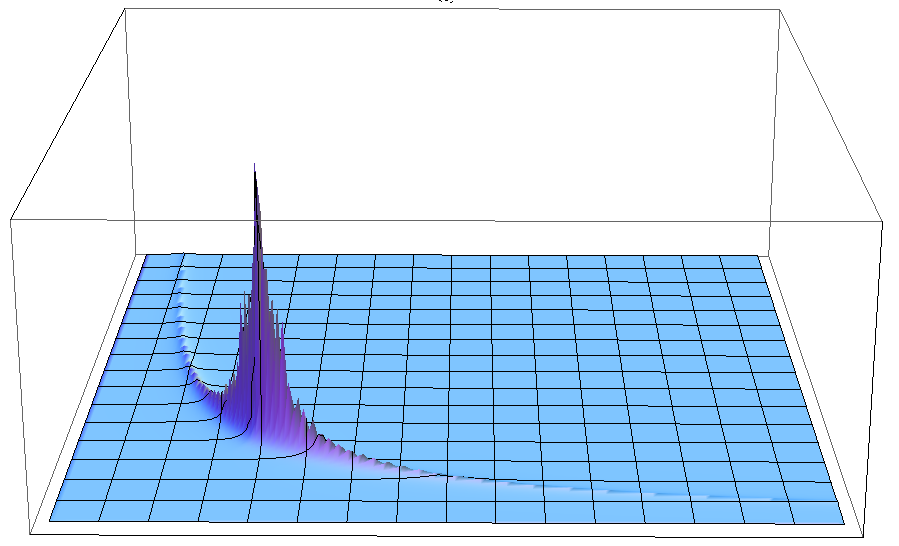}
\caption{Here $e^{-\lambda\tau}=0.09$. Concentration on proper-time hyperbola is now evident. This type of effective ``foliation" is implied by the assumed form of the diffeomorphism. The support of the wavepacket moves towards the light cone. The fine peaks on the plot are an artifact of Wolfram Mathematica algorithm.}
\end{figure}
\begin{figure}
\includegraphics[width=12cm]{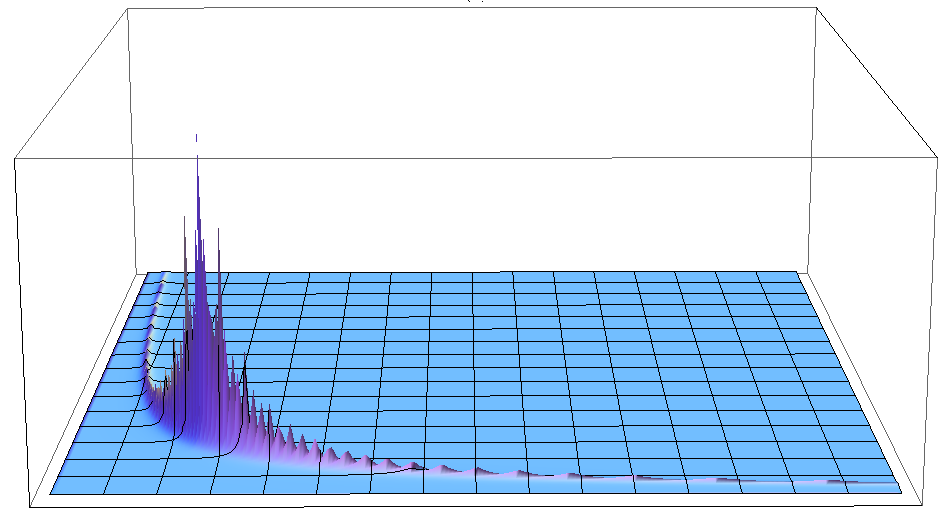}
\caption{The state of the Universe for $e^{-\lambda\tau}=0.04$. As $\tau$ increases towards $+\infty$, the wavepacket approaches the boundary $x_ax^a=0$.}
\end{figure}
The evolution parameter is $\tau$, and although we defined the initial state at $\tau=0$, one could monitor the evolution in $\tau$ backwards towards $-\infty$. The wave packet would then shrink in space but expand in time. Thus, a long time before $\tau=0$ the Universe was localized in a tiny region of space but its timelike extension was enormous.

The next four figures show the dynamics of a wavepacket that is initially two-peaked. The two peaks do not overlap and thus are mutually orthogonal. The dynamics we consider does not have matrix elements between the two orthogonal states, so the state (of our single Universe) remains in a superposition of two non-overlapping parallel universes occupying non-overlapping regions of space-time.
\begin{figure}
\includegraphics[width=12cm]{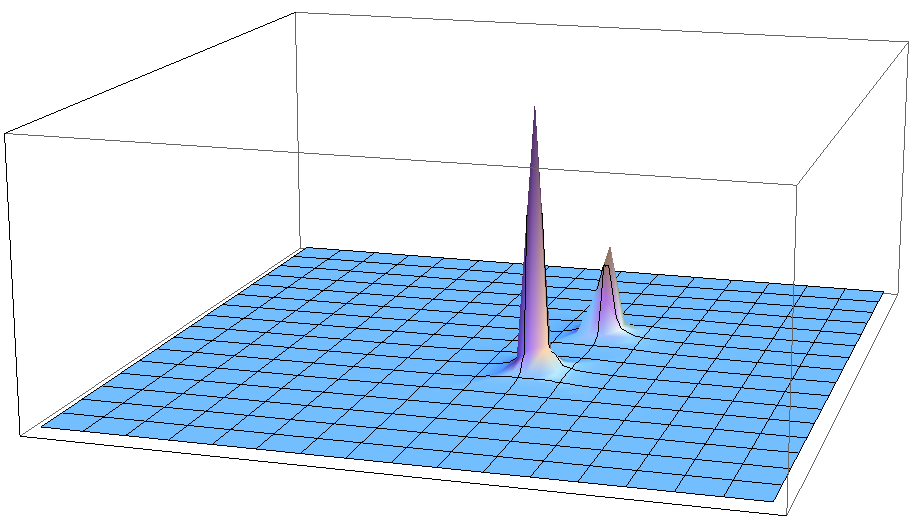}
\caption{``Parallel Universes": The two-peaked initial wavepacket at $\tau=0$.}
\end{figure}
\begin{figure}
\includegraphics[width=12cm]{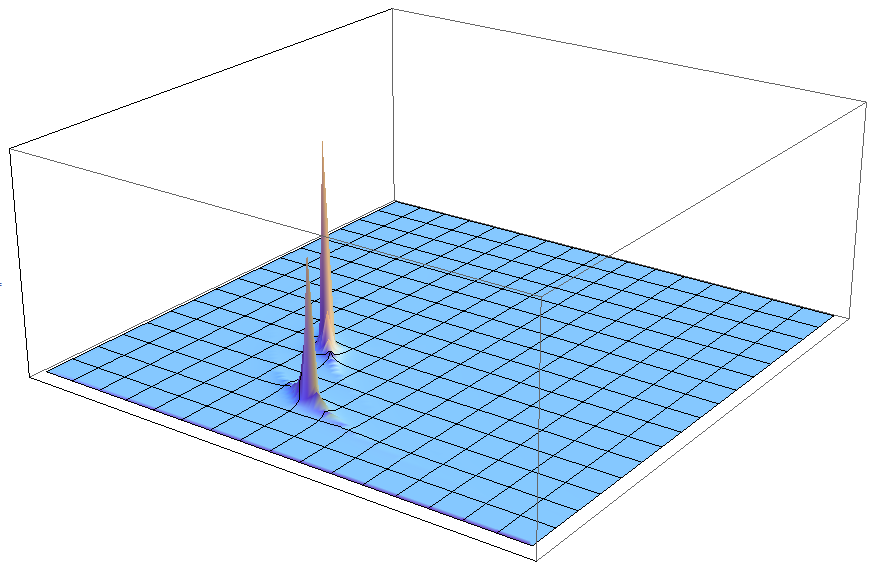}
\caption{Now $e^{-\lambda\tau}=0.3$. The two peaks remain orthogonal.}
\end{figure}
\begin{figure}
\includegraphics[width=12cm]{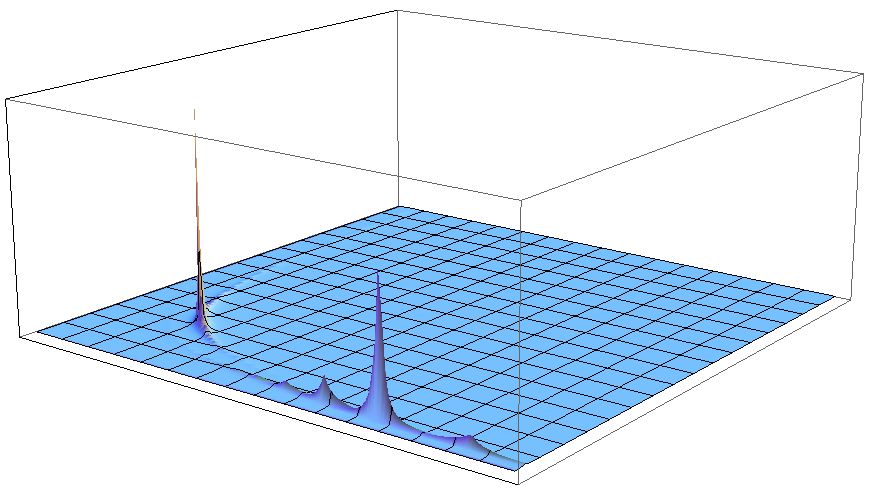}
\caption{Here $e^{-\lambda\tau}=0.1$. Concentration on proper-time hyperbolas is clearly evident.}
\end{figure}
\begin{figure}
\includegraphics[width=12cm]{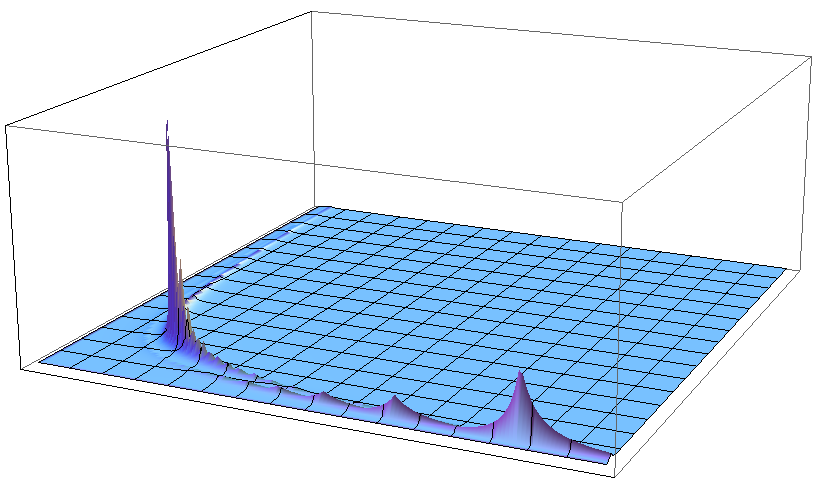}
\caption{The state of the Universe for $e^{-\lambda\tau}=0.07$.}
\end{figure}

\section{R\'enyi generalization}

Distribution of galaxies is known to possess certain multi-fractal properties. Some authors suggest \cite{Jizba} that maximizing Shannon's entropy on a multi-fractal is equivalent to maximizing directly the R\'enyi entropy
\be
\frac{1}{1-q}\ln \left( \sum_j p_j^q\right)
\ee
without invoking the multi-fractal structure explicitly. Therefore, a kind of fractal generalization of the exponential case can be obtained if one replaces Shannon's entropy by R\'enyi's entropy of order $q$. The question is if the remaining averages occurring in the Massieu function should be kept in the same form as in the Shannon case, or maybe one should modify them as well? The answer was proposed by Naudts and one of the present authors in \cite{CN,N}. The key element was to realize that R\'enyi's entropy was originally derived by R\'enyi in \cite{Renyi} by considering the same random variable $\ln(1/p_j)$ as in the Shannon definition, but what had to be changed was the averaging procedure. More concretely, R\'enyi derived his entropy by replacing linear averaging by an appropriate Kolmogorov-Nagumo average. Keeping this in mind, the authors of \cite{CN} defined the R\'enyi Massieu function as an analogous Kolmogorov-Nagumo average of all the random variables, constraints included. When applied to processes such as Zipf-Mandelbrot law in linguistics \cite{Montemurro}, or protein folding dynamics \cite{TBM}, the approach from \cite{CN} directly gave a formula consistent with experimental data. This should be contrasted with the approach based on Tsallis thermodynamics \cite{TBM} which required ad hoc modifications in order to reconstruct experimental data beyond a crude linear fit. Let us adapt the procedure from \cite{CN} to the present context.

The appropriate Massieu function reads
\be
E_{\rm KN}
&=&
\varphi^{-1}\left(\sum_{j=0}^1 p_j\varphi[\ln(1/p_j)]\right)
+
\alpha \varphi^{-1}\left(\sum_{j=0}^1 p_j\varphi(1)\right)
+
\beta\varphi^{-1}\left(\sum_{j=0}^1 p_j\varphi[\lambda j\tau]\right),
\ee
where $\varphi$ is a strictly monotonic function which defines a Kolmogorov-Nagumo average. The R\'enyi entropy corresponds to $\varphi(x)=e^{(1-q)x}$, $\varphi^{-1}(x)=(1-q)^{-1}\ln x$.
The R\'enyi form of $\varphi$ is uniquely determined by the requirement that
\be
\varphi^{-1}
\big(\sum_jp_j\varphi(x_j+C)\big)=C+\varphi^{-1}
\big(\sum_jp_j\varphi(x_j)\big)\label{x+C}
\ee
for a constant $C$ (the proof can be found in \cite{Jizba}). This includes the linear case $\varphi(x)\sim x$, reconstructed in the limit $q\to 1$.

The explicit Massieu function now reads
\be
E_{\rm R}
&=&
\frac{1}{1-q}\ln\left(\sum_{j=0}^1 p_j^q\right)
\nonumber\\
&\pp=&
+
\alpha
+
\alpha \frac{1}{1-q}\ln\left(\sum_{j=0}^1 p_j\right)
+
\beta\frac{1}{1-q}\ln\left(\sum_{j=0}^1 p_je^{(1-q)\lambda j\tau}\right).
\ee
We have to extremize it under the constraint $p_0+p_1=1$. Denote
\be
\overline{\lambda\tau}=\frac{1}{1-q}\ln\left(\sum_{j=0}^1 p_je^{(1-q)\lambda j\tau}\right).
\ee
Computing
\be
\frac{\partial E_{R}}{\partial p_0}
&=&
\frac{1}{1-q}
\left(
\frac{qp_0^{q-1}}{\sum_{j=0}^1 p_j^q}
+
\alpha \frac{1}{\sum_{j=0}^1 p_j}
+
\beta\frac{1}{\sum_{j=0}^1 p_je^{(1-q)\lambda j\tau}}
\right)=0,\label{der1}\\
\frac{\partial E_{R}}{\partial p_1}
&=&
\frac{1}{1-q}
\left(
\frac{qp_1^{q-1}}{\sum_{j=0}^1 p_j^q}
+
\alpha \frac{1}{\sum_{j=0}^1 p_j}
+
\beta\frac{e^{(1-q)\lambda \tau}}{\sum_{j=0}^1 p_je^{(1-q)\lambda j\tau}}
\right)=0\label{der2},
\ee
we obtain two equations with consistency condition
\be
q+\alpha+\beta=0.
\ee
Denoting $\gamma=-\beta/q$, $1-\gamma=-\alpha/q$, and incorporating the constraint, one finds
\be
\frac{p_0^{q-1}}{\sum_{j=0}^1 p_j^q}
&=&
1-\gamma
+
\gamma e^{(q-1)\overline{\lambda\tau}},\\
\frac{p_1^{q-1}}{\sum_{j=0}^1 p_j^q}
&=&
1-\gamma
+
\gamma e^{(q-1)(\overline{\lambda\tau}-\lambda\tau)},
\ee
which leads to the final form
\be
p_1/p_0
&=&
\left[
\frac{
1-\gamma
+
\gamma e^{(q-1)(\overline{\lambda\tau}-\lambda\tau)}
}
{
1-\gamma
+
\gamma e^{(q-1)\overline{\lambda\tau}}
}
\right]^{1/(q-1)}
.
\ee
Note that $\overline{\lambda\tau}$ in principle depends on $\tau$, so $p_1/p_0$ is defined as an implicit function. Nevertheless,
in the limit $q\to 1$ we reconstruct the exponential case
\be
p_1/p_0\to e^{\beta\lambda\tau}
\ee
as expected on the basis of the Shannon limit of R\'enyi entropies.
The parameter $\gamma=-\beta/q$ should not in itself be regarded as a probability (in principle, $\gamma$ can be negative or greater than 1; in the Shannon case we took $\beta=1$).

Defining, as before, the evolution parameter by $\ln (p_1/p_0)$ one arrives at a R\'enyi generalization of the diffeomorphism from the previous section, 
\be
\eta_\tau=
\left[
\frac{
1-\gamma
+
\gamma e^{(q-1)(\overline{\lambda\tau}-\lambda\tau)}
}
{
1-\gamma
+
\gamma e^{(q-1)\overline{\lambda\tau}}
}
\right]^{1/(q-1)}
\eta.
\ee
Similarly to the usual exponent, the above generalization can be directly obtained from a differential equation.
Indeed, in the previous section we have started with $\eta_\tau=e^{\lambda \tau}\eta$, $\xi_\tau=e^{-\lambda \tau}\xi$, that is with
\be
\frac{d(\eta_\tau/\eta_0)}{d\tau} &=& \lambda \eta_\tau/\eta_0,\label{q=1}\\
\eta_\tau\xi_\tau &=& \eta_0\xi_0\label{q=1a}.
\ee
Let us generalize (\ref{q=1}) to
\be
\frac{d(\eta_\tau/\eta_0)}{d\tau} &=& \lambda_r (\eta_\tau/\eta_0)^r+(\lambda_p-\lambda_r) (\eta_\tau/\eta_0)^p ,\label{q}
\ee
but keep (\ref{q=1a}) unchanged. (\ref{q}) was introduced by Tsallis, Bemski and Mendes \cite{TBM} as a model of protein re-association dynamics, and later employed by Montemurro \cite{Montemurro} in quantitative linguistics. Comparison with both protein and linguistic data showed that a very good fitting could be obtained for $r=1$ and an appropriate $p\neq 1$. A yet better fitting was found if both $r$ and $p$ where different from 1. An analogous two-parameter generalization was derived in \cite{CN} directly from Kolmogorov-Nagumo averages. For $r=1$ one gets a special case of the Bernoulli equation,
\be
\frac{d(\eta_\tau/\eta_0)}{d\tau} &=& \lambda_1 \eta_\tau/\eta_0+(\lambda_p-\lambda_1) (\eta_\tau/\eta_0)^p ,\label{q,r=1}
\ee
which can be solved with arbitrary initial condition at $\tau_0$. The result is
\be
\eta_{\tau}
&=&
\left[
\frac{1-\frac{\lambda_p }{\lambda_1}+\frac{\lambda_p }{\lambda_1}e^{(1-p)\lambda_1\tau}}
{1-\frac{\lambda_p }{\lambda_1}+\frac{\lambda_p }{\lambda_1}e^{(1-p)\lambda_1\tau_0}}
\right]^{1/(1-p)}
\eta_{\tau_0}\\
&=&
\epsilon_{\tau,\tau_0}\eta_{\tau_0},\label{sol q}
\ee
and has the form we have derived from the Massieu function $E_{\rm R}$ if $1-p=q-1$,

The two-time function $\epsilon_{\tau,\tau_0}=\epsilon_{\tau}/\epsilon_{\tau_0}$, $\epsilon_{\tau}=\epsilon_{\tau,0}$, satisfies the groupoid composition property
\be
\epsilon_{\tau_1,\tau_2}\epsilon_{\tau_2,\tau_3}=\epsilon_{\tau_1,\tau_3}.\label{grupoid}
\ee
Asymptotically, for large $\tau$, one finds $\eta_\tau\approx (\lambda_p/\lambda_1)^{\frac{1}{1-p}}e^{\lambda_1 \tau}\eta_0$, and for small $\tau$
\be
\eta_\tau\approx
\eta_0\Big[1+\lambda_p(1-p)\tau\Big]^{\frac{1}{1-p}}.
\ee
The dynamical system (\ref{q}) has a nontrivial covariance property under changes of scale, $\epsilon_\tau\to a\,\epsilon_\tau$, $a\in \mathbb{R}_+$, $da/d\tau=0$,
\be
\frac{d(a\epsilon_\tau)}{d\tau}
&=&
\lambda_r a^{1-r}(a\epsilon_\tau)^r+(\lambda_p-\lambda_r)a^{1-p} (a\epsilon_\tau)^p
\nonumber\\
&=& \lambda_r'(a\epsilon_\tau)^r+(\lambda_p'-\lambda_r')(a\epsilon_\tau)^p.
\ee
Solving $\lambda_r'=\lambda_r a^{1-r}$, $\lambda_p'-\lambda_r'= (\lambda_p-\lambda_r)a^{1-p}$, we obtain a matrix representation
\be
\left(
\begin{array}{c}
\lambda_p'\\
\lambda_r'
\end{array}
\right)
&=&
\left(
\begin{array}{ccc}
a^{1-p} & ,& a^{1-r}-a^{1-p}\\
0 & , &a^{1-r}
\end{array}
\right)
\left(
\begin{array}{c}
\lambda_p\\
\lambda_r
\end{array}
\right)
=
T_{p,r}(a)\left(
\begin{array}{c}
\lambda_p\\
\lambda_r
\end{array}
\right),
\ee
$T_{p,r}(a)T_{p,r}(b)=T_{p,r}(ab)$, of the multiplicative group $\mathbb{R}_+$. The exponential case corresponds to the trivial representation with $p=r=1$.

The small-$\tau$ regime then corresponds to the case
$e^{\lambda_1(1-p)\tau}\approx 1+\lambda_1(1-p)\tau$ which coincides with the well known Tsallis result relating his entropy with measures of Lyapunov instability \cite{Tsallis}. From the Kolmogorov-Nagumo-R\'enyi perspective the maximal entropy results of Tsallis may be regarded as linear approximations to the more exact models based on R\'enyi entropies and nonlinear averaging.

Now let us check the evolution of $R_\tau$ implied by (\ref{sol q}):
\be
R_\tau^2
&=&
\langle U_\tau \psi|\hat r_v^2 U_\tau \psi\rangle
=
\int_0^\infty d\eta\int_{-\infty}^\infty d\xi\,2\eta\xi^2\big|U_\tau f_v(s,\xi)\big|^2
\nonumber\\
&=&
\int_0^\infty d\eta\int_{-\infty}^\infty d\xi\,2\eta\xi^2\big|f_v(\epsilon_\tau\eta,\xi/\epsilon_\tau)\big|^2
=
\epsilon_\tau\int_0^\infty d\eta'\int_{-\infty}^\infty d\xi'\,2\eta'\xi'^2\big|f_v(s',\xi')\big|^2
\nonumber\\
&=&
\epsilon_\tau R^2.
\ee
Accordingly, a multi-crossover generalization of the Hubble law is then given by $R_\tau=\sqrt{\epsilon_\tau}R_0$.

In the next section we discuss the structure of the generator of evolution corresponding to a general $\epsilon_\tau$.

\section{Schr\"odinger equation}

The wave function
\be
U_\tau f_v(\eta,\xi)
&=&
f_v(\eta\epsilon_\tau,
\xi/\epsilon_\tau)
\ee
satisfies
\be
\frac{d}{d\tau}U_\tau f_v(\eta,\xi)
&=&
\frac{d\eta_\tau}{d\tau}\frac{\partial f_v(\eta_\tau,\xi_\tau)}{\partial \eta_\tau}
+
\frac{d\xi_\tau}{d\tau}\frac{\partial f_v(\eta_\tau,\xi_\tau)}{\partial \xi_\tau}
\nonumber\\
&=&
\frac{d\ln\epsilon_\tau}{d\tau}
\Big(
\eta\frac{\partial}{\partial \eta}
-
\xi\frac{\partial }{\partial \xi}
\Big)
U_\tau f_v(\eta,\xi)=-i\hat \omega_\tau\, U_\tau f_v(\eta,\xi).
\ee
The generator of evolution
\be
\hat \omega_\tau &=& \frac{d\ln\epsilon_\tau}{d\tau}
\Big(
\eta\, i\frac{\partial}{\partial \eta}
-
\xi\,i\frac{\partial }{\partial \xi}
\Big)\\
&=& \frac{1}{2}\frac{d\ln\epsilon_\tau}{d\tau}
\Big(
\eta\, p_\eta+p_\eta\eta
-
\xi\,p_\xi-p_\xi\xi
\Big)
\ee
with
\be
p_\eta &=& i\frac{\partial}{\partial \eta},\\
p_\xi &=& i\frac{\partial }{\partial \xi},
\ee
is in general $\tau$-dependent. Still, since $[\hat \omega_\tau,\hat \omega_{\tau'}]=0$, we can integrate the dynamics and arrive at
\be
U_\tau &=& \exp\left(-i\int_0^\tau d\tau'\,\hat \omega_{\tau'}\right)
= e^{\big(
\eta\frac{\partial}{\partial \eta}
-
\xi\frac{\partial }{\partial \xi}
\big)\ln\epsilon_\tau}
= \epsilon_\tau^{
\eta\frac{\partial}{\partial \eta}
-
\xi\frac{\partial }{\partial \xi}}.
\label{U_tau}
\ee
In effect, we have obtained a standard-looking unitary dynamics $U_\tau=e^{-i\varepsilon\hat\Phi}$ with time-{\it independent\/} generator
\be
\hat \Phi &=& \frac{1}{2}\Big(
\eta\, p_\eta+p_\eta\eta
-
\xi\,p_\xi-p_\xi\xi
\Big)
\ee
if we reinterpret $\varepsilon=\ln \epsilon_\tau$ as a new dimensionless evolution parameter (a similar dimensionless evolution parameter occurs in scale relativity \cite{Scale1}). In consequence, in addition to unitarity we obtain a conserved ``average energy"
\be
\Phi=\langle\psi|U_\tau^\dag\hat\Phi U_\tau|\psi\rangle.
\ee
Note that $\epsilon_\tau = \eta_\tau/\eta_0=\eta_\tau/\eta$
is independent of $\eta$ and $\xi$, so the operator in the exponent commutes with $\eta_\tau/\eta$. Therefore,
\be
U_{\tau_2,\tau_1}
&=&
\left(\frac{\eta_{\tau_2}}{\eta_{\tau_1}}\right)^{
\eta\frac{\partial}{\partial \eta}
-
\xi\frac{\partial }{\partial \xi}}
=
\left(\frac{\xi_{\tau_1}}{\xi_{\tau_2}}\right)^{
\eta\frac{\partial}{\partial \eta}
-
\xi\frac{\partial }{\partial \xi}}
=
\left(\frac{\eta_{\tau_2}}{\eta_{\tau_1}}\right)^{
\eta\frac{\partial}{\partial \eta}
}
\left(\frac{\xi_{\tau_1}}{\xi_{\tau_2}}\right)^{
-
\xi\frac{\partial }{\partial \xi}}
,\label{eta/eta}
\ee
with $U_{\tau}=U_{\tau,0}$. The relation (\ref{eta/eta}) between the diffeomorphism $(\eta,\xi)\to (\eta_\tau,\xi_\tau)$ and the unitary transformation $U_\tau$ is very simple.
The composition law
\be
U_{\tau_3,\tau_2}U_{\tau_2,\tau_1}=U_{\tau_3,\tau_1}
\ee
follows immediately from (\ref{eta/eta}).

Let us check the action of $U_{\tau_2,\tau_1}$ on monomials,
\be
U_{\tau_2,\tau_1}\eta_\tau^n
&=&
\frac{\eta_{\tau_2}^n\eta_\tau^n}{\eta_{\tau_1}^n},
\nonumber\\
U_{\tau_2,\tau_1}\xi_\tau^n
&=&
\frac{\xi_{\tau_2}^n\xi_\tau^n}{\xi_{\tau_1}^n}.
\ee
So,
\be
U_{\tau_2,\tau_1}\eta_{\tau_1}^n &=& \eta_{\tau_2}^n,\label{Ueta}\\
U_{\tau_2,\tau_1}\xi_{\tau_1}^n &=& \xi_{\tau_2}^n,\label{Uxi}
\ee
and for any $f(\eta_{\tau_1},\xi_{\tau_1})$ which can be expanded in a power series one finds
\be
U_{\tau_2,\tau_1}f(\eta_{\tau_1},\xi_{\tau_1}) = f(\eta_{\tau_2},\xi_{\tau_2})
\ee
as required.

Now let us return to the `fractal' evolution parameter
\be
\varepsilon
&=&
\frac{1}{1-p}\ln
\left(1-\frac{\lambda_p}{\lambda_1}+\frac{\lambda_p}{\lambda_1}e^{\lambda_1(1-p)\tau}\right)
\\
&=&
\varphi^{-1}
\left(1-\frac{\lambda_p}{\lambda_1}+\frac{\lambda_p}{\lambda_1}\varphi(\lambda_1\tau)\right)
,
\ee
where $\varphi(x)=e^{(1-p)x}$, $\varphi^{-1}(x)=(1-p)^{-1}\ln x$ is again the Kolmogorov-Nagumo function employed by R\'enyi in his derivation of generalized entropies.
For $0\leq \lambda_p\leq \lambda_1$ the parameters $\lambda_p/\lambda_1$ and  $1-\lambda_p/\lambda_1$ are probabilities and the Kolmogorov-Nagumo average is indeed an average. However, condition (\ref{x+C}) holds even for $\lambda_p/\lambda_1$ and  $1-\lambda_p/\lambda_1$ non-interpretable as probabilities, provided
(\ref{x+C}) is defined. Actually, in the entropic derivation of the generalized exponent we encountered $\gamma$ and $1-\gamma$ that could be negative or greater than 1. From our perspective this means one can also consider the case $\lambda_p> \lambda_1$, but then $\tau$ cannot be arbitrary (see the next section).

\section{The case $1-p=q-1=1$}

The simplest and yet quite close to the expected form of the Bernoulli dynamics is the case $p=0$. It corresponds to the R\'enyi entropy of order $q=2$. To begin with, note that the function
\be
\epsilon_\tau
&=&
1-\lambda_0/\lambda_1+e^{\lambda_1 \tau}\lambda_0/\lambda_1
\ee
leads to exponential expansion
\be
R_\tau
&=&
R_0\sqrt{1-\lambda_0/\lambda_1+e^{\lambda_1 \tau}\lambda_0/\lambda_1}
\nonumber\\
&\approx& R_0\sqrt{\lambda_0/\lambda_1}e^{\lambda_1 \tau/2}
\ee
for large $\tau$, and a square-root law
\be
R_\tau
&=&
R_0\sqrt{1-\lambda_0/\lambda_1+e^{\lambda_1 \tau}\lambda_0/\lambda_1}
\nonumber\\
&\approx& R_0\sqrt{1+\lambda_0 \tau}
\ee
for small $\tau$. We assume $\lambda_0>0$, $\lambda_1>0$. For $\tau\to-\infty$ one finds
\be
R_{-\infty}
&=&
R_0\sqrt{1-\lambda_0/\lambda_1}
\ee
which suggests $\lambda_0<\lambda_1$. However, one expects that as long as $e^{\lambda_1 \tau}$ can be approximated by $1+\lambda_1 \tau$ (hence for small values of $\lambda_1 \tau$), the dynamics is of a square-root type $R_\tau\sim \sqrt{\tau}$, a fact meaning $\lambda_0 \tau\gg 1$. Putting these two conditions together we conclude that in the crossover regime one finds $\lambda_0 \tau\gg 1$ and $\lambda_1 \tau\ll 1$. We therefore have to investigate also the case $\lambda_0\gg\lambda_1$. This leads us to the critical value $\tau_0$,
\be
1-\lambda_0/\lambda_1+e^{\lambda_1 \tau_0}\lambda_0/\lambda_1=0.
\ee
In such a case there exists an absolute origin of the dynamics
\be
\tau_0=\frac{1}{\lambda_1}\ln \frac{\lambda_0-\lambda_1}{\lambda_0}<0
\ee
corresponding to $R_{\tau_0}=0$ and $\ln \epsilon_{\tau_0}=-\infty$. Note that at $\tau_0$ the entire Universe is localized on the line $X^a=v^a s$, $0<s<\infty$. In this way the fiducial world line is no longer arbitrary, but is defined by the support of the initial condition $U_{\tau_0}\psi(x^0,x^1)$.

The existence of two evolution parameters, $\tau$ and $\varepsilon=\ln\epsilon_\tau$, leads to a kind of paradox. Namely, for $\lambda_0>\lambda_1$ the asymptotic properties of $\epsilon_\tau$ imply that the evolution operator $e^{-i\varepsilon \hat\Phi}$ involves an effective evolution parameter which is an arbitrary real number, $-\infty<\varepsilon<\infty$. So, from the point of view of $e^{-i\varepsilon \hat\Phi}$ the dynamics looks as if the system evolved in time from $-\infty$ till ``now", but from the point of view of $\tau$ the evolution starts at a finite $\tau_0$. On the other hand, for $\lambda_0<\lambda_1$ the parameter $\tau$ takes any real value, $-\infty<\tau<\infty$, but $\epsilon_{-\infty}>0$ and thus $\varepsilon =\ln \epsilon_{-\infty}$ is finite. The evolution then looks as if the system existed since a finite time $\varepsilon$, and yet $\tau$ is unlimited from below. Of course, these remarks apply to any $q$, not only to $q=2$.

\section{Evolution operator in space-time variables}

The analysis given in the preceding sections heavily relied on covariant coordinates $\eta$ and $\xi$, which are not completely natural if one switches to higher dimensions. So, from the point of view of higher-dimensional generalizations it is important to rephrase the results in terms of space-time variables $x^a$. In order to do so, we begin with (\ref{Ueta})--(\ref{Uxi}), implying
\be
U_\tau
\left(
\begin{array}{c}
x^0\\
x^1
\end{array}
\right)
&=&
\sqrt{\epsilon_\tau}s
\left(
\begin{array}{cc}
\cosh\Xi & \sinh\Xi\\
\sinh\Xi & \cosh\Xi
\end{array}
\right)
\left(
\begin{array}{c}
\cosh(\xi/\epsilon_\tau)\\
\sinh(\xi/\epsilon_\tau)
\end{array}
\right)
\nonumber\\
&=&
\frac{\sqrt{\epsilon_\tau}}{2}s^{1-1/\epsilon_\tau}
\left(
\begin{array}{cc}
\cosh\Xi & \sinh\Xi\\
\sinh\Xi & \cosh\Xi
\end{array}
\right)
\left(
\begin{array}{c}
(x^0+x^1)^{1/\epsilon_\tau}+(x^0-x^1)^{1/\epsilon_\tau}\\
(x^0+x^1)^{1/\epsilon_\tau}-(x^0-x^1)^{1/\epsilon_\tau}
\end{array}
\right).
\nonumber
\ee
Let us change variables from $(\eta,\xi)$ to $(x^0, x^1)$
\be
\eta\frac{\partial}{\partial \eta}
-
\xi\frac{\partial }{\partial \xi}
&=&
\frac{1}{2}x^a\partial_a
-
\xi(x)
\big(
x^1\partial_0
+
x^0\partial_1
\big)\nonumber.
\ee
The second term involves a generator of a representation of a Lorentz transformation: If $y_a=\Lambda{_a}{^b}x_b$ then
\be
\psi(y^0,y^1)
&=&
e^{\zeta_{ab} (x^a\partial^b-x^b\partial^a)/2}\psi(x^0,x^1)
,
\ee
where $\zeta_{01}=-\zeta_{10}=\zeta$.
However, in spite of this, the whole term $-\xi(x^1\partial_0+x^0\partial_1)$ does {\it not\/} generate the Lorentz transformation $f(x)\mapsto f(X)$ since $\xi$ depends on $x$.
The problem is similar to that with the other term, $x^a\partial_a$. It involves the generator of translations $\partial_a$, but $x^a\partial_a$ generates rescalings
\be
e^{\lambda x^a\partial_a}f(x) = f(e^\lambda x)
\ee
and not translations,
\be
e^{\lambda y^a\partial_a}f(x) = f(x+\lambda y),
\ee
occurring only for $y^a$ independent of $x$.
So, denote $L=-\xi(x^1\partial_0+x^0\partial_1)$, $D=x^a\partial_a$.
The commutator $[L,D]=0$ vanishes since $L$ preserves homogeneity of functions $f(x^0,x^1)$ and $D$ is the Euler homogeneity operator.

The dynamics is given by
\be
U_{\tau_2,\tau_1}\psi_{\tau_1}(x)=\psi_{\tau_2}(x)
\ee
where
\be
U_{\tau_2,\tau_1}
&=&
\left(\frac{x_{\tau_2}^2}{x_{\tau_1}^2}\right)^{D/2+L}
\nonumber\\
&=&
e^{(\varepsilon_2-\varepsilon_1)D/2}e^{(\varepsilon_2-\varepsilon_1)L}
.
\ee
The `fractal' parameters are defined by
\be
\varepsilon_j=\ln(x_{\tau_j}^2)=\ln \left[\phi_{\tau_j}(x)^2\right],\quad j=1,2.
\ee
Since
$x_{\tau_2}^2/x_{\tau_1}^2=\epsilon_{\tau_2}/\epsilon_{\tau_1}$ is, by construction, independent of $x^a$ it thus commutes with $D$ and $L$.

One can weaken the latter condition. Indeed, $x_{\tau_2}^2/x_{\tau_1}^2=\phi_{\tau_2}(x)^2/\phi_{\tau_1}(x)^2$ commutes with $D$ and $L$ if $\phi_{\tau}$ is 1-homogeneous, $\phi_{\tau}(\lambda x)=\lambda \phi_{\tau}(x)$.
In order to generalize the form of $U_{\tau_2,\tau_1}$ to $1+3$ dimensions consider a Lorentz transformation $\Lambda(x)$ which maps $x^a$ into some fiducial point $X^a$. There exist parameters $\xi_{ab}(x)$ and generators $S^{ab}$ of SO(1,3) such that $\Lambda(x)=\exp (\xi_{ab}(x)S^{ab}/2)$. Now consider two points $x^a$ and $y^a$ related by a general Lorentz transformation $\Lambda=\exp (\zeta_{ab}S^{ab}/2)$, $y_a=\Lambda{_a}{^b}x_b$, and let
$L^{ab}=x^a\partial^b-x^b\partial^a$ be the generator of
$f(y)=e^{\zeta_{ab}L^{ab}/2}f(x)$. Then the $1+3$ dimensional analog of the $1+1$ dimensional $e^L$ is
$e^{\xi_{ab}(x)L^{ab}/2}$. Let us note that the operator $e^{\xi_{ab}(x)L^{ab}/2}$ is not uniquely defined by $X^a$ and $x^a$ since the Lorentz transformation $x^a\mapsto X^a$ is defined up to little groups of $X^a$ and $x^a$. Our choice of the background space-time $V_+$ implies that $X^a$ and $x^a$ are time-like for a finite $\tau$. The little group is thus O(3) or SU(2).

\section{Homogeneity and isotropy: Space-time vs. Universe}

Our model space-time is homogeneous and isotropic, but the Universe is neither homogeneous nor isotropic, at least not exactly. The point is that the ``shape" of the Universe is defined by the wave packet $\psi(x^0,x^1)$. Such a wave packet  can be as close as possible to a uniform distribution, but cannot be everywhere constant, of course. This is a general feature of our formalism and cannot be eliminated. Fortunately, various inhomogeneities and anisotropies are in fact observed in the Universe we live in, so the property seems physically acceptable. The more subtle point is that we rescale the variable $\xi$ with respect to a fixed fiducial reference frame defined by $v_a$. The presence of $v_a$ apparently breaks uniformity of proper-time $s$-hyperbolas. One can think of this symmetry breaking in two ways. First of all, we have shown that in models that in a finite time approach a zero-volume state characterized by $R_{\tau_0}=0$ the support of the wave function coincides with the world line $x_a(s)=v_a s$. So this ``preferred" world line is encoded in the initial choice of the wave function of the Universe, and we are back to the problem of non-uniformity of $\psi(x^0,x^1)$. Secondly, the status of $v_a$ is similar to that of the point $x_a=0$ in Minkowski space. Indeed, in standard Poincar\'e group we have two subgroups: 4-translations and Lorentz transformations. The Lorentz transformations are equivalent to hyperbolic and ordinary rotations around a preferred but arbitrary $x_a=0$. This arbitrariness of $x_a=0$ is controlled by the 4-translation subgroup. So the Poincar\'e group controls two kinds of arbitrariness: The one of location of the ``origin", and the one of the frame attached to this ``origin".

In our case we encounter a similar logical structure. The ``origin" is controlled by Lorentz transformations $v_a\to v_a'=\Lambda{_a}{^b}v_a$, so the Lorentz group plays here a role analogous to the 4-translations subgroup of the Poincar\'e group. In fact, Lorentz transformations act in the $\xi$-space as translations. But the analogue of the Lorentz subgroup of the Poincar\'e group is, in our formalism, the rescaling $\xi\to\xi/\epsilon$. Thus, the dynamical group behind our dynamics is the semidirect product of SO(1,3) and changes of scale on its homogeneous space.

\section{Conclusions}

The problem of combining groups of diffeomorphisms originating from some classical geometric theory with unitary dynamics of the Universe can be formulated in a way which resembles Hilbert-space approaches of Koopman \cite{K} and von Neumann \cite{vN}, proposed in 1930s  in the context of classical mechanics. The resulting dynamics of the ``wave function of the Universe" possesses features analogous to expansion known from realistic models of cosmology, including crossovers from a ``radiation dominated" $\sqrt{\tau}$ phase, to the ``dark energy" accelerating expansion for large $\tau$. This type of dynamics follows from the assumption that the flow of time follows from an extremal entropy principle for R\'enyi entropies. The model does not employ a preferred time-foliation but nevertheless a kind of effective foliation occurs in a dynamical way as a consequence of the form of the diffeomorphism that defines the dynamics. In the explicit examples discussed in the paper the effective foliation converges towards spacelike  hyperbolas which subsequently asymptotically evolve into a light-cone.

The resulting picture of an evolving Universe is different from the usual one where it is space-time itself which expands. In our approach space-time is an arena for evolution of the Universe, the latter being identified with the region of space-time occupied by the wavepacket. So we have a dynamical Universe evolving in a static space-time. In all the examples we have concentrated on a $1+1$ dimensional space-time since all homogeneous isotropic space-times are effectively mathematically $1+1$ dimensional. Nevertheless, the full theory must be formulated in at least $1+3$ dimensions, and only at such a stage one can think of comparison with exact observational cosmology. This final step has not been performed in the paper, but the unitary dynamics was brought to a form that does not crucially depend on $1+1$ dimensionality of the formalism, and is easy to generalize to higher dimensions and more general fields.

\end{document}